\documentclass[11pt]{article}
\usepackage[centertags]{amsmath}
\usepackage{amsfonts}
\usepackage{amssymb}
\usepackage{amsthm}
\usepackage{newlfont}
\usepackage{graphicx}
\usepackage{multirow}
\usepackage{rotating}

\theoremstyle{definition}

\theoremstyle{remark}

\theoremstyle{theorem}

\begin{document}
\thispagestyle{empty}

%
%

\title{ Numerical approximation of modified non-linear SIR model of computer viruses  }
\author{ Samad Noeiaghdam \footnote{Corresponding author, E-mail addresses: s.noeiaghdam.sci@iauctb.ac.ir; samadnoeiaghdam@gmail.com
~~~~~Tel.: +98 9143527552}
 }
  \date{}
 \maketitle

\begin{center}
\scriptsize{ Department of Mathematics, Central Tehran Branch, Islamic Azad University, Tehran, Iran. \\
}
\end{center}
\begin{abstract}
In this paper, the non-linear modified epidemiological model of
computer viruses is illustrated. For this aim, two semi-analytical
methods, the differential transform method (DTM) and the
Laplace-Adomian decomposition method (LADM) are applied. The
numerical results are estimated for different values of iterations
and compared to the results of the LADM and the homotopy analysis
transform method (HATM). Also, graphs of residual errors and phase
portraits of approximate solutions for $n=5,10,15$ are demonstrated.
The numerical approximations show the performance of the LADM in
comparison to the LADM and the HATM.

 \vspace{.5cm}{\it keywords: Non-linear
Susceptible-Infected-Recovered model, Differential transform
 method, Laplace transformations, Adomian decomposition method.}
\end{abstract}
\section{Introduction}
The computer viruses are malware programs that have been able to
infect thousands of computers and have hurt billions dollar in
computers around the world. The virus should never be considered to
be harmless and remain in the system. There are several types of
viruses that can be categorized according to their source,
technique, file type that infects, where they are hiding, the type
of damage they enter, the type of operating system, or the design on
which they are attacking. We can introduce some of famous and
malicious viruses such as ILOVEYOU, Melissa, My Doom, Code Red,
Sasser, Stuxnet and so on. Therefore, it is important that we study
the methods to analyze, track, model, and protect against viruses.
In recent years, many scientists have been illustrated the
epidemiological models of computer viruses
\cite{1,5,13,11,3,4,7,8,9,12,6}. These models have been estimated by
many mathematical methods such as collocation method \cite{18},
 homotopy analysis method \cite{20,man5}, variational iteration method
\cite{man2} and others \cite{4,17}.

One of applicable and important models is the classical
Susceptible-Infected-Recovered (SIR) computer virus propagation
model \cite{11,16,4} which is presented in the following form:
\begin{equation}\label{1}
\begin{array}{l}
\displaystyle  \frac{dS(t)}{dt} = f_1 - \lambda S(t) I(t) - d S(t),   \\
    \\
\displaystyle   \frac{dI(t)}{dt} = f_2 +  \lambda S(t) I(t) - \varepsilon I(t) - d R(t),  \\
    \\
\displaystyle    \frac{dR(t)}{dt} = f_3 + \varepsilon I(t) - d R(t), \\
\end{array}
\end{equation}
where
\begin{equation}\label{2}
  S(0) = S_0(t),
  I(0) = I_0(t),
  R(0) = R_0(t),
\end{equation}
are the initial conditions of non-linear system of Eqs. (\ref{1}).
Functions and initial values of system (\ref{1}) are given in Table
\ref{t1}.

\begin{table}[h]
\caption{ List of parameters and functions. }\label{t1}
 \centering
\scalebox{0.9}{
\begin{tabular}{|c|l|c|}
\hline
 Parameters &  ~~~~~~~~~~~~~~~~~~~~~~~~~Meaning   & Values      \\
 $\&$ Functions &&\\
    \hline
$S(t)$& Susceptible computers at time $t$&$S(0)=20$\\
$I(t)$& Infected computers at time $t$ &$I(0)=15$\\
$R(t)$& Recovered computers at time $t$ &$R(0)=10$\\
$f_1, f_2, f_3$& Rate of external computers connected to the network &$f_1=f_2=f_3=0$\\
$\lambda$&  Rate of infecting for  susceptible computer &$\lambda=0.001$\\
$\varepsilon$& Rate of recovery for infected computers &$\varepsilon=0.1$\\
$d$& Rate of removing from the network &$d=0.1$\\
 \hline
 \end{tabular}
 }
\end{table}

Recently, several numerical and semi-analytical methods are
introduced to solve the mathematical and engineering problems
\cite{me1,me2,me3,man3,man4,man6,me99} that we can apply them to
solve the non-linear model (\ref{1}). The DTM and the LADM are two
important and efficient tools to solve the linear and non-linear
problems arising in the mathematics, physics and engineering
\cite{fa1,fa2,18dtm,15dtm,16dtm,17dtm,waz1}. Specially, the LADM
\cite{26,25,27} obtained  by combining the Adomian decomposition
method \cite{ab4,waz2,waz3,ab3,ab2,waz4} and the Laplace
transformations \cite{ab1} similar to the HATM
\cite{21,man3,man5,22}, Laplace homotopy perturbation method
\cite{29,28,30} and so on.

The aim of this paper is to apply the DTM and the LADM to find the
approximate solution of non-linear epidemiological system of Eqs.
(\ref{1}). The numerical results are compared with the HATM
\cite{21,man3,man5,22} by plotting the residual errors function for
different iterations. Also, the phase portraits of approximate
solutions for $n=10$ and different functions of $S(t), I(t)$ and
$R(t)$ are presented. The numerical results show the abilities and
capabilities of the LADM in comparison to the DTM and the HATM.

\section{Differential transform method}

Transformation of the $k$-th derivative of a function in one
variable is as follows
\begin{equation}\label{3d}
F(k) = \frac{1}{k!} \left[\frac{d^k f(t)}{dt^k}\right]_{t=t_0},
\end{equation}
and the inverse transformation is defined by
\begin{equation}\label{4d}
f(t) = \sum_{k=0}^{\infty} F(k) (t-t_0)^k
 \end{equation}
where $F(k)$ is the differential transform of $f(t)$. In actual
applications, the function $f(t)$ is expressed by a finite series
and Eq. (\ref{4d}) can be rewritten as follows:
\begin{equation}\label{5d}
f(t) = \sum_{k=0}^{N} F(k) (t-t_0)^k
 \end{equation}
where $N$ is decided by the convergence of natural frequency. The
fundamental operations of DTM have been given in Table \ref{t2}.

\begin{table}[h]
\caption{ Main operations of DTM. }\label{t2}
 \centering
\scalebox{0.9}{
\begin{tabular}{|c|c|}
\hline Original functions & Transformed functions\\
  \hline
&  \\
$ f(t) = u(t) \pm v(t) $& $F(k)= U(k) \pm V(k)$\\
& \\
 $f(t) = \beta u(t)$& $F(k)= \beta U(k)$\\
& \\
 $f(t) = u(t) v(t)$ & $F(k)= \sum_{i=0}^k U(k) V_{k-s}(k)$\\
& \\
$ f(t) = \frac{d u(t)}{dt}$& $F(k)=(k+1)U(k+1)$\\
& \\
 $f(t) =\frac{d^m u(t)}{dt^m} $&$ F(k)= (k+1)(k+2) \cdots (k+m) U(k+m) $\\
& \\
$ f(t) = \int_{t_0}^t u(\xi) d\xi $& $F(k)=\frac{U(k-1)}{k}, k \geq 1$\\
& \\
$ f(t) = t^m $&$ F(k)= \delta(k-m) $\\
& \\
$ f(t) = \exp (\lambda t) $&$ F(k)= \frac{\lambda^k }{k!}$\\
& \\
 $f(t) = \sin (\omega t + \alpha)$&$ F(k)= \frac{\omega^k }{k!} \sin (\frac{\pi k}{2} + \alpha)$\\
& \\
 $f(t) = \cos (\omega t + \alpha)$&$ F(k)= \frac{\omega^k }{k!} \cos (\frac{\pi k}{2} + \alpha) $\\
& \\
 \hline
 \end{tabular}
 }
\end{table}

By applying the presented method to system of Eqs. (\ref{1}), we get
\begin{equation}\label{6d}
 \begin{array}{l}
 \displaystyle   S_{k+1} = \frac{1}{k+1} \left[ f_1 - \lambda \sum_{i=0}^k S_i I_{k-i} - d S_k   \right], \\
     \\
\displaystyle   I_{k+1} = \frac{1}{k+1} \left[ f_2 - \lambda \sum_{i=0}^k S_i I_{k-i} - \varepsilon I_k - d R_k   \right], \\
     \\
 \displaystyle  R_{k+1} = \frac{1}{k+1} \bigg[ f_3 - \varepsilon  I_{k} - d R_k   \bigg]. \\
 \end{array}
\end{equation}
The differential transform method series solution for system
(\ref{1}) can be obtained as
\begin{equation}\label{7}
\begin{array}{l}
\displaystyle   S(t) = \sum_{j=0}^n S_jt^j, \\
    \\
\displaystyle   I(t) = \sum_{j=0}^n I_jt^j, \\
    \\
\displaystyle   R(t) = \sum_{j=0}^n R_jt^j, \\
\end{array}
\end{equation}

\section{Laplace-Adomian decomposition method}

We apply the Laplace transformation $\mathcal{L}$ as
\begin{equation}\label{3}
\begin{array}{l}
\displaystyle  \mathcal{L} [S(t)] = \frac{S(0)}{z} + \frac{\mathcal{L} [f_1]}{z} - \frac{\lambda}{z} \mathcal{L} [ S(t) I(t)] - \frac{d}{z} \mathcal{L} [S(t)],   \\
    \\
\displaystyle   \mathcal{L} [I(t)]  =  \frac{I(0)}{z} + \frac{\mathcal{L} [f_2]}{z} +  \frac{\lambda}{z}  \mathcal{L} [S(t) I(t)] - \frac{\varepsilon}{z} \mathcal{L} [I(t)]  - \frac{d}{z} \mathcal{L} [R(t)],  \\
    \\
\displaystyle    \mathcal{L} [R(t)] = \frac{R(0)}{z} + \frac{\mathcal{L} [f_3]}{z} + \frac{\varepsilon}{z} \mathcal{L} [I(t)] - \frac{d}{z} \mathcal{L} [R(t)]. \\
\end{array}
\end{equation}

By putting the initial conditions we have
\begin{equation}\label{4}
\begin{array}{l}
\displaystyle  \mathcal{L} [S(t)] = \frac{S(0)}{z} + \frac{f_1}{z^2} - \frac{\lambda}{z} \mathcal{L} [ A] - \frac{d}{z} \mathcal{L} [S(t)],   \\
    \\
\displaystyle   \mathcal{L} [I(t)]  =  \frac{I(0)}{z} + \frac{f_2}{z^2}  +  \frac{\lambda}{z}  \mathcal{L} [A] - \frac{\varepsilon}{z} \mathcal{L} [I(t)]  - \frac{d}{z} \mathcal{L} [R(t)],  \\
    \\
\displaystyle    \mathcal{L} [R(t)] = \frac{R(0)}{z} + \frac{f_3}{z^2}  + \frac{\varepsilon}{z} \mathcal{L} [I(t)] - \frac{d}{z} \mathcal{L} [R(t)], \\
\end{array}
\end{equation}
where $A=SI$ and
\begin{equation}\label{5}
S= \sum_{j=0}^\infty S_j,~~~~~ I= \sum_{j=0}^\infty I_j,~~~~~ R=
\sum_{j=0}^\infty R_j.
\end{equation}
Also, the non-linear operator $A$  is called the Adomian polynomials
and it is presented as
\begin{equation}\label{6}
A= \sum_{j=0}^\infty A_j,
\end{equation}
where
\begin{equation}\label{7}
\begin{array}{l}
  A_0=S_0 I_0, \\
    \\
  A_1= S_0I_1+S_1 I_0, \\
    \\
  A_2 = S_0 I_2 + S_1I_1 + S_2I_0, \\
  \\
  A_3 = S_0I_3+S_1I_2+S_2I_1 +S_3I_0,\\

  \\
  A_4 = S_0I_4+ S_1 I_3 + S_2I_2 + S_3 I_1 + S_4 I_0, \\
~~~  \vdots
\end{array}
\end{equation}

By substituting series (\ref{5}) and (\ref{6}) into (\ref{4}) we get
\begin{equation}\label{8}
\begin{array}{l}
\displaystyle  \mathcal{L} \left[\sum_{j=0}^\infty S_j \right] = \frac{S(0)}{z} + \frac{f_1}{z^2} - \frac{\lambda}{z} \mathcal{L} \left[ \sum_{j=0}^\infty A_j \right] - \frac{d}{z} \mathcal{L} \left[\sum_{j=0}^\infty S_j \right],   \\
    \\
\displaystyle   \mathcal{L} \left[\sum_{j=0}^\infty I_j \right]  =  \frac{I(0)}{z} + \frac{f_2}{z^2}  +  \frac{\lambda}{z}  \mathcal{L} \left[\sum_{j=0}^\infty A_j \right] - \frac{\varepsilon}{z} \mathcal{L} \left[\sum_{j=0}^\infty I_j \right]  - \frac{d}{z} \mathcal{L} \left[\sum_{j=0}^\infty R_j \right],  \\
    \\
\displaystyle    \mathcal{L} \left[\sum_{j=0}^\infty R_j \right] = \frac{R(0)}{z} + \frac{f_3}{z^2}  + \frac{\varepsilon}{z} \mathcal{L} \left[\sum_{j=0}^\infty I_j \right] - \frac{d}{z} \mathcal{L} \left[\sum_{j=0}^\infty R_j \right]. \\
\end{array}
\end{equation}
Now, the following relations can be obtained:
\begin{equation}\label{9}
 \begin{array}{l}
\displaystyle  \mathcal{L} [S_0] =   \frac{S(0)}{z} + \frac{f_1}{z^2},  \\
\\
\displaystyle \mathcal{L} [I_0] =   \frac{I(0)}{z} + \frac{f_2}{z^2},  \\
\\
\displaystyle \mathcal{L} [R_0] =   \frac{R(0)}{z} + \frac{f_3}{z^2},  \\
\\
\displaystyle \mathcal{L} [S_1] = \frac{\lambda}{z} \mathcal{L} \left[ A_0 \right] - \frac{d}{z} \mathcal{L} \left[S_0 \right],  \\
 \\
\displaystyle \mathcal{L} [I_1] =  \frac{\lambda}{z}  \mathcal{L} \left[A_0 \right] - \frac{\varepsilon}{z} \mathcal{L} \left[I_0 \right]  - \frac{d}{z} \mathcal{L} \left[R_0 \right],  \\
 \\
\displaystyle  \mathcal{L} [R_1] =  \frac{\varepsilon}{z}
\mathcal{L} \left[I_0 \right] - \frac{d}{z} \mathcal{L} \left[R_0
\right],
 \end{array}
\end{equation}
and for term $j$ we have
\begin{equation}\label{10}
\begin{array}{l}
\displaystyle \mathcal{L} [S_j] = \frac{\lambda}{z} \mathcal{L} \left[ A_{j-1} \right] - \frac{d}{z} \mathcal{L} \left[S_{j-1} \right],  \\
 \\
\displaystyle \mathcal{L} [I_j] =  \frac{\lambda}{z}  \mathcal{L} \left[A_{j-1} \right] - \frac{\varepsilon}{z} \mathcal{L} \left[I_{j-1} \right]  - \frac{d}{z} \mathcal{L} \left[R_{j-1} \right],  \\
 \\
\displaystyle  \mathcal{L} [R_j] =  \frac{\varepsilon}{z}
\mathcal{L} \left[I_{j-1} \right] - \frac{d}{z} \mathcal{L}
\left[R_{j-1} \right].
\end{array}
\end{equation}

Applying the inverse Laplace transformation $ \mathcal{L}^{-1}$ for
first equations of (\ref{9}) as  follows
\begin{equation}\label{11}
\begin{array}{l}
  S_0 = S(0) + f_1 t, \\
  \\
  I_0 = I(0) + f_2 t,\\
  \\
  R_0= R(0) + f_3 t,
\end{array}
\end{equation}
By putting $S_0, I_0, R_0$ in second equations of (\ref{9}) and
using the Laplace transformations we have
\begin{equation}\label{12}
\begin{array}{ll}
\displaystyle \mathcal{L} [S_1] &\displaystyle = \frac{\lambda}{z}\left(\frac{S(0) I(0)}{z} + \frac{S(0) f_2}{z^2}+\frac{I(0) f_1}{z^2}+ \frac{2 f_1 f_2}{z^3}\right) - \frac{d}{z} \left(\frac{S(0)}{z}+\frac{f_1}{z^2}\right),  \\
 \\
\displaystyle \mathcal{L} [I_1] &\displaystyle =  \frac{\lambda}{z}  \left(\frac{S(0) I(0)}{z} + \frac{S(0) f_2}{z^2}+\frac{I(0) f_1}{z^2}+ \frac{2 f_1 f_2}{z^3}\right) - \frac{\varepsilon}{z} \left(\frac{I(0)}{z}+\frac{f_2}{z^2}\right) \\
\\
&\displaystyle  - \frac{d}{z} \left(\frac{R(0)}{z}+\frac{f_3}{z^2}\right),  \\
 \\
\displaystyle  \mathcal{L} [R_1] & \displaystyle =
\frac{\varepsilon}{z} \left(\frac{I(0)}{z}+\frac{f_2}{z^2}\right) -
\frac{d}{z} \left(\frac{R(0)}{z}+\frac{f_3}{z^2}\right),
 \end{array}
\end{equation}
and by applying the inverse Laplace transform $ \mathcal{L}^{-1}$ we
can find $S_1, I_1$ and $R_1$. By repeating above process, the other
terms $S_2, \cdots, S_j, I_2, \cdots, I_j, R_2, \cdots, R_j$ can be
obtained. By using the relations
\begin{equation}\label{n-app}
S_n= \sum_{j=0}^n S_j,~~~~~ I_n= \sum_{j=0}^n I_j,~~~~~ R_n=
\sum_{j=0}^n R_j,
\end{equation}
the $n$-th order approximate solutions can be estimated.

\section{Numerical Illustration}
In this section, the numerical results of the DTM and the LADM for
solving the system of Eqs. (\ref{1}) are presented. The approximate
solutions for $n=5$ by using the DTM are obtained in the following
form
$$
\begin{array}{l}
S_{5}(t)=  20 - 2.3 t + 0.15425 t^2 - 0.00790458 t^3 + 0.000309711 t^4, \\
  \\
I_{5}(t)= 15 - 2.2 t + 0.04575 t^2 + 0.00573792 t^3 - 0.000406169 t^4,\\
\\
R_{5}(t)= 10 + 0.5 t - 0.135 t^2 + 0.006025 t^3 - 7.17708\times
10^{-6} t^4,
\end{array}
$$
and for $n=10$ we have
$$
\begin{array}{l}
S_{10}(t)= 20 - 2.3 t + 0.15425 t^2 - 0.00790458 t^3 + 0.000309711
t^4\\
\\
~~~~~~~~~~ -7.74864\times 10^{-6} t^5 - 1.35996\times 10^{-8} t^6 +
1.41005\times 10^{-8}
t^7\\
\\
~~~~~~~~~~ - 8.93373\times 10^{-10} t^8 + 3.32927\times 10^{-11} t^9,\\
 \\
I_{10}(t)=15 - 2.2 t + 0.04575 t^2 + 0.00573792 t^3 - 0.000406169
t^4 \\
\\~~~~~~~~~~+
 9.82135\times
10^{-6} t^5 + 1.12052\times 10^{-7} t^6 - 1.97453\times 10^{-8}
t^7\\
\\
~~~~~~~~~~+
 9.96904\times
10^{-10} t^8 - 3.2067\times
10^{-11} t^9,\\
 \\
R_{10}(t)=10 + 0.5 t - 0.135 t^2 + 0.006025 t^3 - 7.17708\times
10^{-6} t^4 \\
\\~~~~~~~~~~ -
 7.97984\times
10^{-6} t^5 + 2.96686\times 10^{-7} t^6 - 2.63764\times 10^{-9}
t^7\\
\\
~~~~~~~~~~ -
 2.13846\times
10^{-10} t^8 + 1.34528\times 10^{-11} t^9,
\end{array}
$$
and finally for $n=15$ the approximate solutions are obtained as
$$
\begin{array}{l}
S_{15}(t)=20 - 2.3 t + 0.15425 t^2 - 0.00790458 t^3 + 0.000309711
t^4 \\
\\~~~~~~~~~~ -
 7.74864\times
10^{-6} t^5 - 1.35996\times 10^{-8} t^6 + 1.41005\times 10^{-8}
t^7\\
\\
~~~~~~~~~~-
 8.93373\times
10^{-10} t^8 + 3.32927\times 10^{-11} t^9 - 5.68453\times 10^{-13}
t^{10}\\
\\
~~~~~~~~~~ -2.34166\times 10^{-14} t^{11} + 2.59234\times 10^{-15}
t^{12} - 1.28786\times 10^{-16} t^{13} +
 3.78868\times
10^{-18} t^{14},\\
 \\
I_{15}(t)=15 - 2.2 t + 0.04575 t^2 + 0.00573792 t^3 - 0.000406169
t^4 \\
\\~~~~~~~~~~ +
 9.82135\times
10^{-6} t^5 + 1.12052\times 10^{-7} t^6 - 1.97453\times 10^{-8}
t^7\\
\\~~~~~~~~~~
+
 9.96904\times
10^{-10} t^8 - 3.2067\times 10^{-11} t^9 + 4.21668\times 10^{-13}
t^{10} \\
\\~~~~~~~~~~+
 2.88892\times
10^{-14} t^{11} - 2.70438\times 10^{-15} t^{12} + 1.28307\times
10^{-16} t^{13} -
 3.62709\times
10^{-18} t^{14},\\
 \\
R_{15}(t)= 10 + 0.5 t - 0.135 t^2 + 0.006025 t^3 - 7.17708\times
10^{-6} t^4\\
\\~~~~~~~~~~ -
 7.97984\times
10^{-6} t^5 + 2.96686\times 10^{-7} t^6 - 2.63764\times 10^{-9}
t^7\\
\\~~~~~~~~~~
-
 2.13846\times
10^{-10} t^8 + 1.34528\times 10^{-11} t^9 - 4.55198\times 10^{-13}
t^{10}\\
\\~~~~~~~~~~ +
 7.97151\times
10^{-15} t^{11} + 1.74314\times 10^{-16} t^{12} - 2.21438\times
10^{-17} t^{13} +
 1.07465\times
10^{-18} t^{14}.
\end{array}
$$


Now, by applying the LADM we get
$$
\begin{array}{lll}
  S_0(t)= 20,&
  I_0(t) =15, &
  R_0(t) =10, \\
  \\
    S_1(t)=-2.3 t, &
  I_1(t) = -2.2 t,&
  R_1(t) = 0.5 t,\\
  \\
    S_2(t)=0.15425 t^2, &
  I_2(t) = 0.04575 t^2,&
  R_2(t) = -0.135 t^2,\\
  \\
  ~~~~~~~~\vdots& ~~~~~~~~\vdots& ~~~~~~~~\vdots\\
  \\
    S_{10}(t)=-5.68453\times 10^{-13} t^{10}, &
  I_{10}(t) = 4.21668\times 10^{-13} t^{10}, &
  R_{10}(t) = -4.55198\times 10^{-13} t^{10},\\
  \\
    ~~~~~~~~\vdots& ~~~~~~~~\vdots& ~~~~~~~~\vdots
\end{array}
$$
and finally the approximate solution of epidemiological model of
computer viruses (\ref{1}) for $n=10$ is in the following form
$$
\begin{array}{l}
\displaystyle  S_{10}(t)= \sum_{j=0}^{10} S_j(t) = 20 - 2.3 t +
0.15425 t^2 - 0.00790458 t^3 + 0.000309711 t^4  \\
 \\
~~~~~~~~~~~~~~~~~~~~~~~ -
 7.74864\times 10^{-6} t^5 - 1.35996\times 10^{-8} t^6 + 1.41005\times
10^{-8} t^7 \\
\\
~~~~~~~~~~~~~~~~~~~~~~~ - 8.93373\times 10^{-10} t^8 + 3.32927\times 10^{-11} t^9 - 5.68453\times 10^{-13} t^{10}, \\
    \\
\displaystyle  I_{10} (t)= \sum_{j=0}^{10} I_j (t)= 15 - 2.2 t +
0.04575 t^2 + 0.00573792 t^3 - 0.000406169 t^4 \\
\\
~~~~~~~~~~~~~~~~~~~~~~~ +
 9.82135\times 10^{-6} t^5
  + 1.12052\times 10^{-7} t^6 - 1.97453\times 10^{-8} t^7 \\
  \\
  ~~~~~~~~~~~~~~~~~~~~~~~+
 9.96904\times 10^{-10} t^8 - 3.2067\times 10^{-11} t^9 + 4.21668\times 10^{-13} t^{10}, \\
\\
\displaystyle  R_{10}(t)= \sum_{j=0}^{10} R_j(t) = 10 + 0.5 t -
0.135 t^2 + 0.006025 t^3 - 7.17708\times 10^{-6} t^4 \\
\\
~~~~~~~~~~~~~~~~~~~~~~~ - 7.97984\times 10^{-6} t^5
 + 2.96686\times 10^{-7} t^6 - 2.63764\times 10^{-9} t^7 \\
 \\
 ~~~~~~~~~~~~~~~~~~~~~~~ -
2.13846\times 10^{-10} t^8 + 1.34528\times 10^{-11} t^9 -
4.55198\times 10^{-13} t^{10}.
\end{array}
$$

In order to show the accuracy of the presented methods, following
residual errors are presented. Also, the numerical results are
compared to the obtained results of the HATM for $n=5,10$. The
results are presented in Tables \ref{t2}, \ref{t3} and \ref{t4}.
\begin{equation}\label{13}
\begin{array}{l}
\displaystyle  E_{n,S}(t) = S'_n(t) - f_1 + \lambda S_n(t) I_n(t) + d S_n(t), \\
    \\
\displaystyle   E_{n,I}(t) = I'_n(t) - f_2 -  \lambda S_n(t) I_n(t) + \varepsilon I_n(t) + d R_n(t),  \\
    \\
\displaystyle   E_{n,R}(t) = R'_n(t) - f_3 - \varepsilon I_n(t) + d R_n(t). \\
\end{array}
\end{equation}

\begin{figure}
\centering
$$\begin{array}{c}
 \includegraphics[width=2.5in]{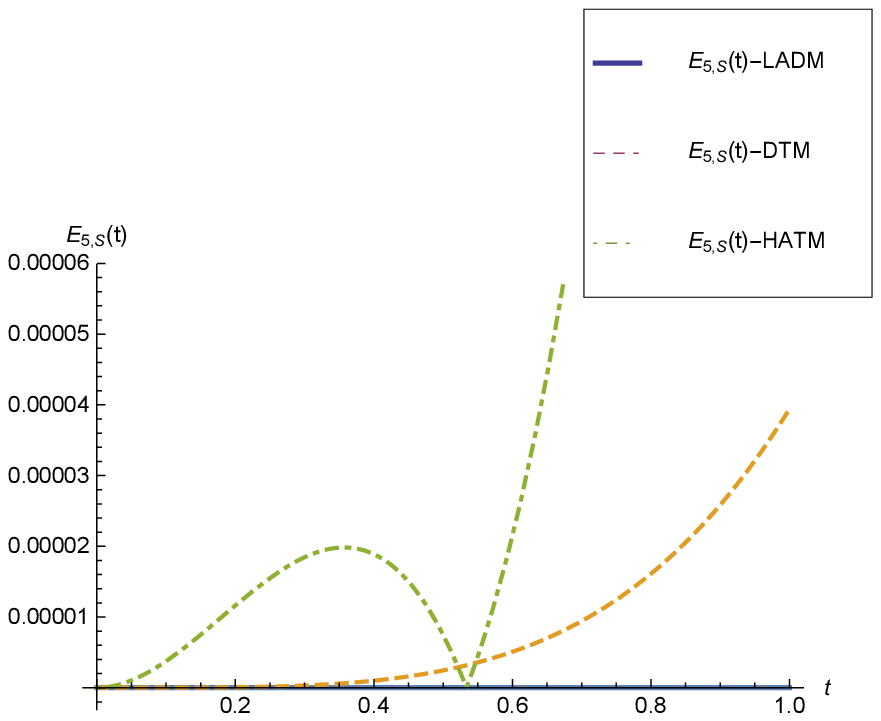}\\
 \\
 \includegraphics[width=2.5in]{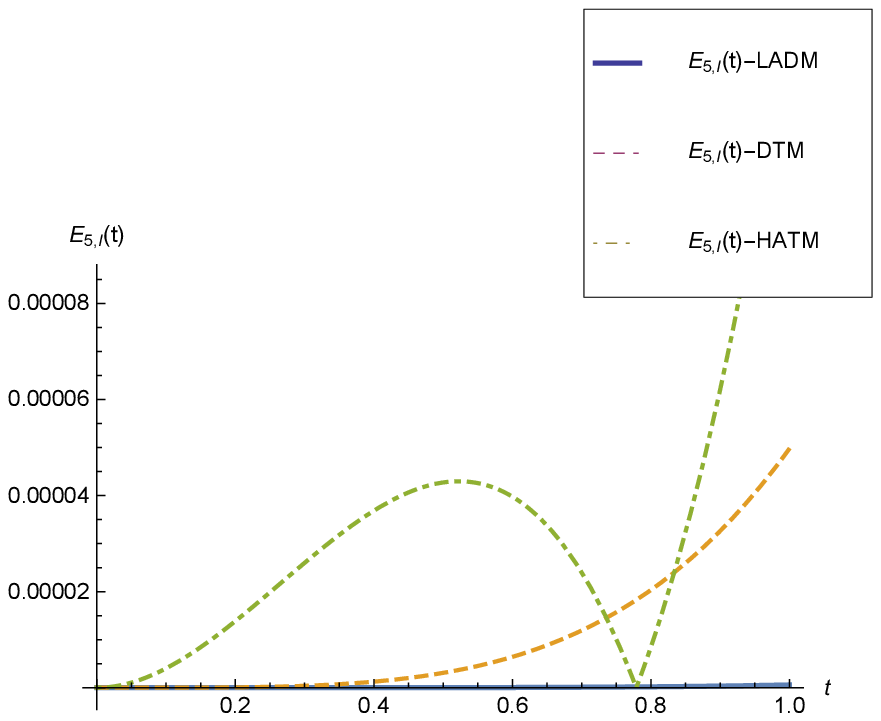}\\
 \\
 \includegraphics[width=2.5in]{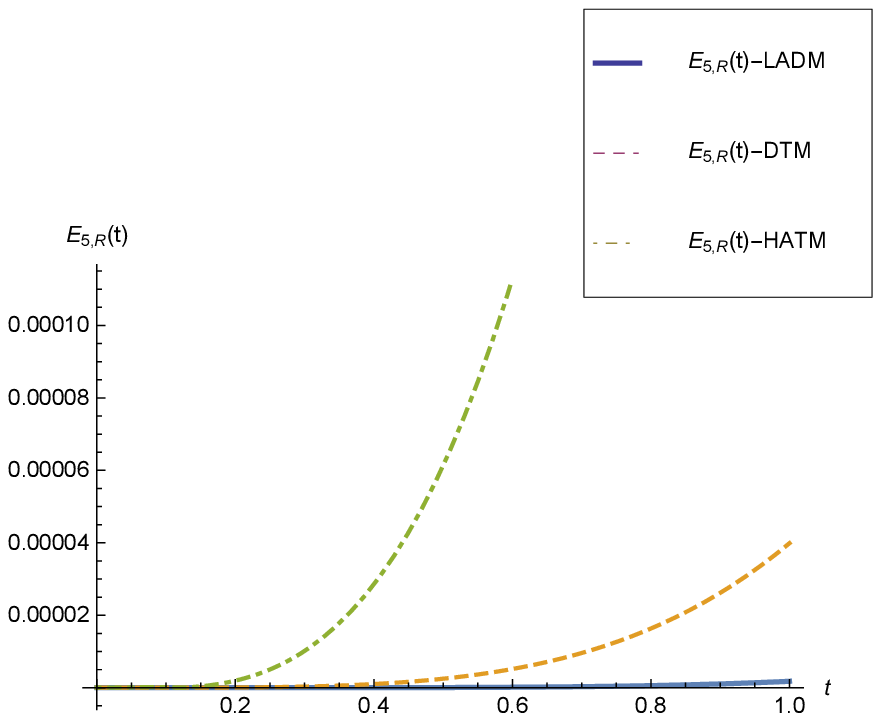}
\end{array}$$
  \caption{Comparison between error functions of LADM, DTM and HATM for $S_{5}(t), I_{5}(t), R_{5}(t)$. }\label{f1}
\end{figure}

\begin{figure}
\centering
$$\begin{array}{c}
 \includegraphics[width=2.5in]{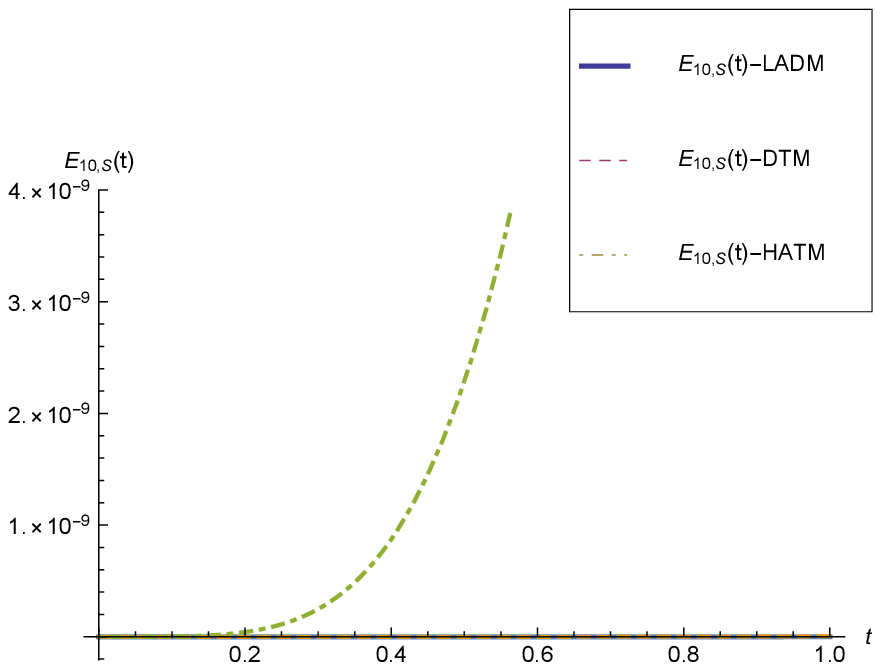}\\
 \\
 \includegraphics[width=2.5in]{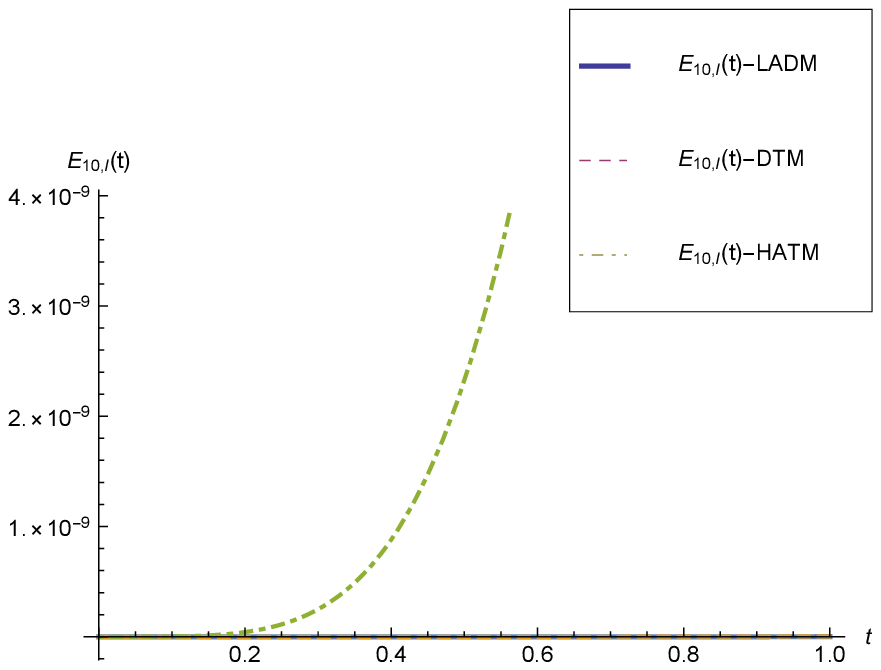}\\
 \\
 \includegraphics[width=2.5in]{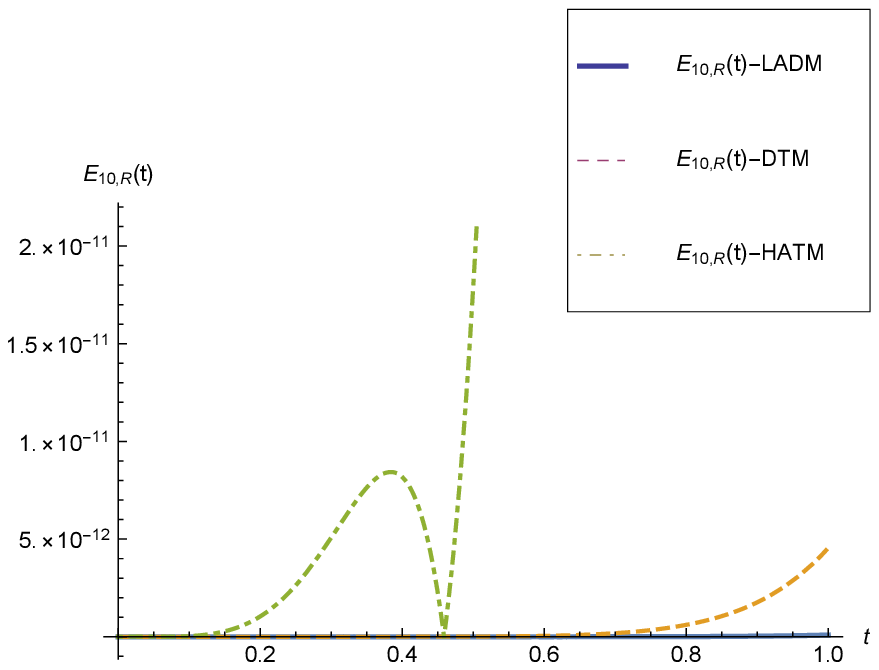}
\end{array}$$
  \caption{Comparison between error functions of LADM, DTM and HATM for $S_{10}(t), I_{10}(t), R_{10}(t)$. }\label{f2}
\end{figure}

\begin{figure}
\centering
$$\begin{array}{c}
 \includegraphics[width=2.5in]{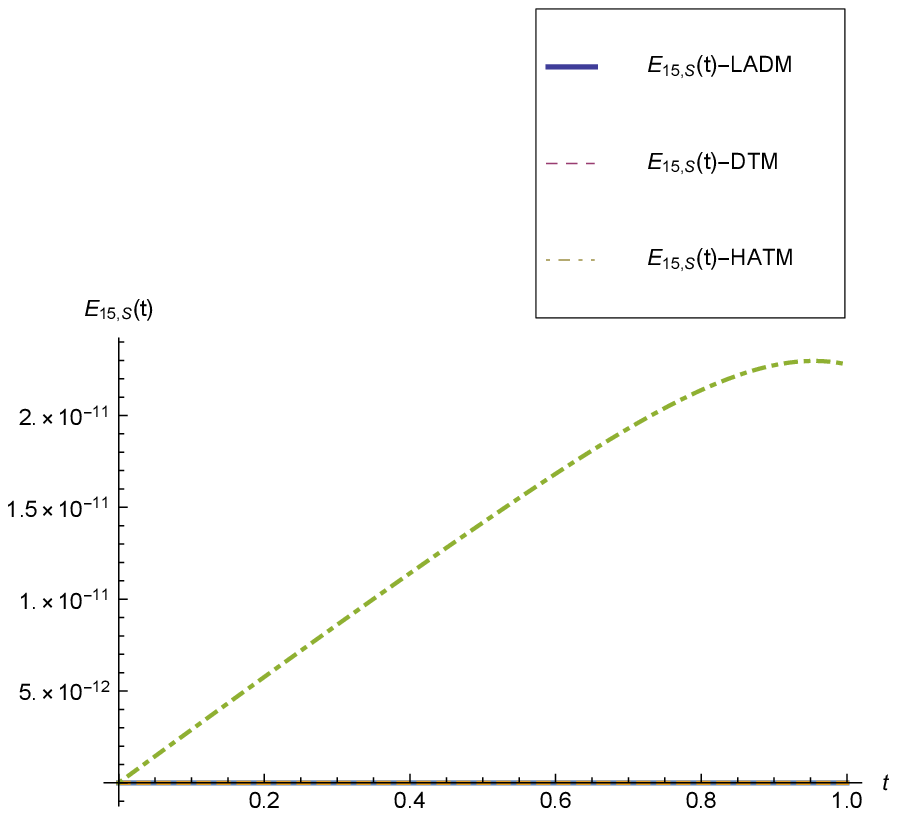}\\
 \\
 \includegraphics[width=2.5in]{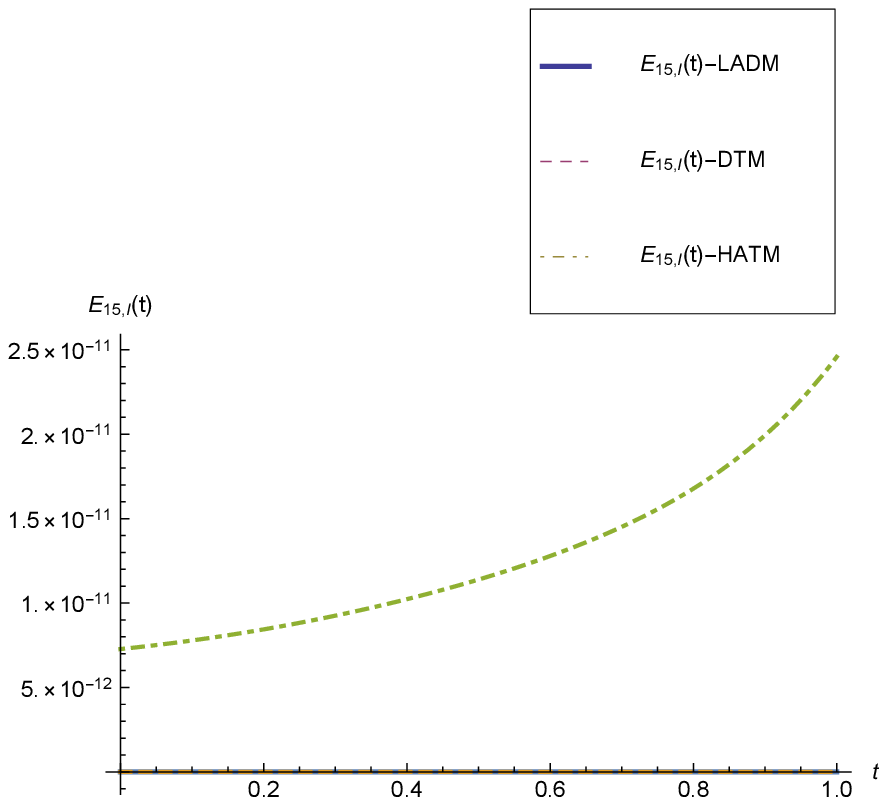}\\
 \\
 \includegraphics[width=2.5in]{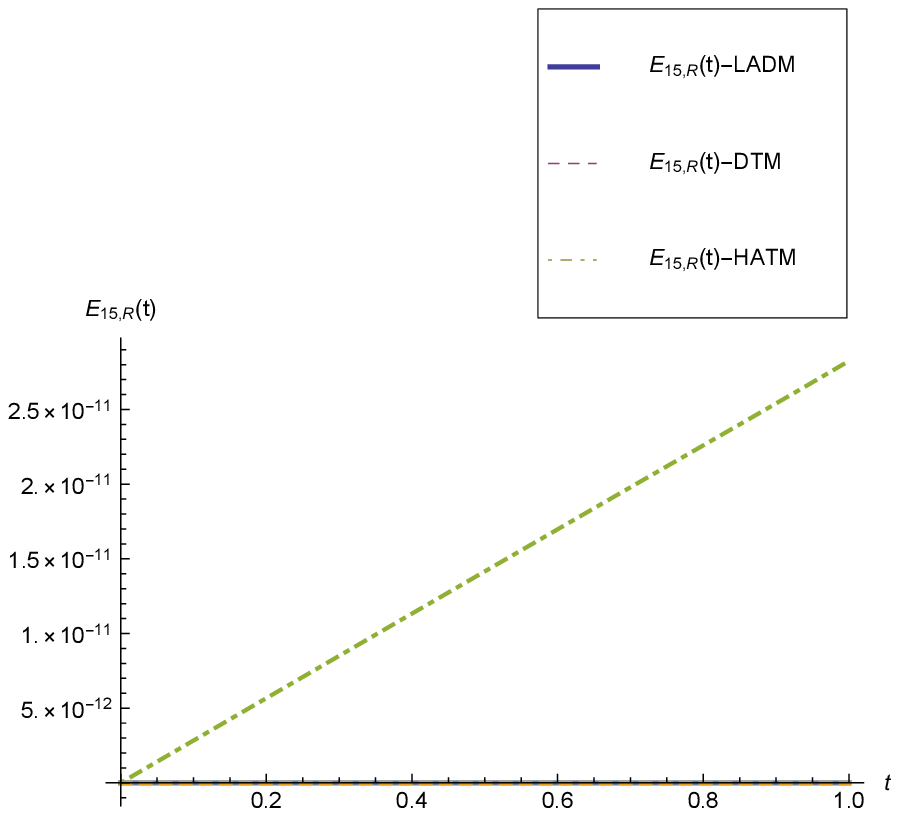}
\end{array}$$
  \caption{Comparison between error functions of LADM, DTM and HATM for $S_{15}(t), I_{15}(t), R_{15}(t)$.}\label{f3}
\end{figure}

\begin{figure}
\centering
$$\begin{array}{ccc}
 \includegraphics[width=2in]{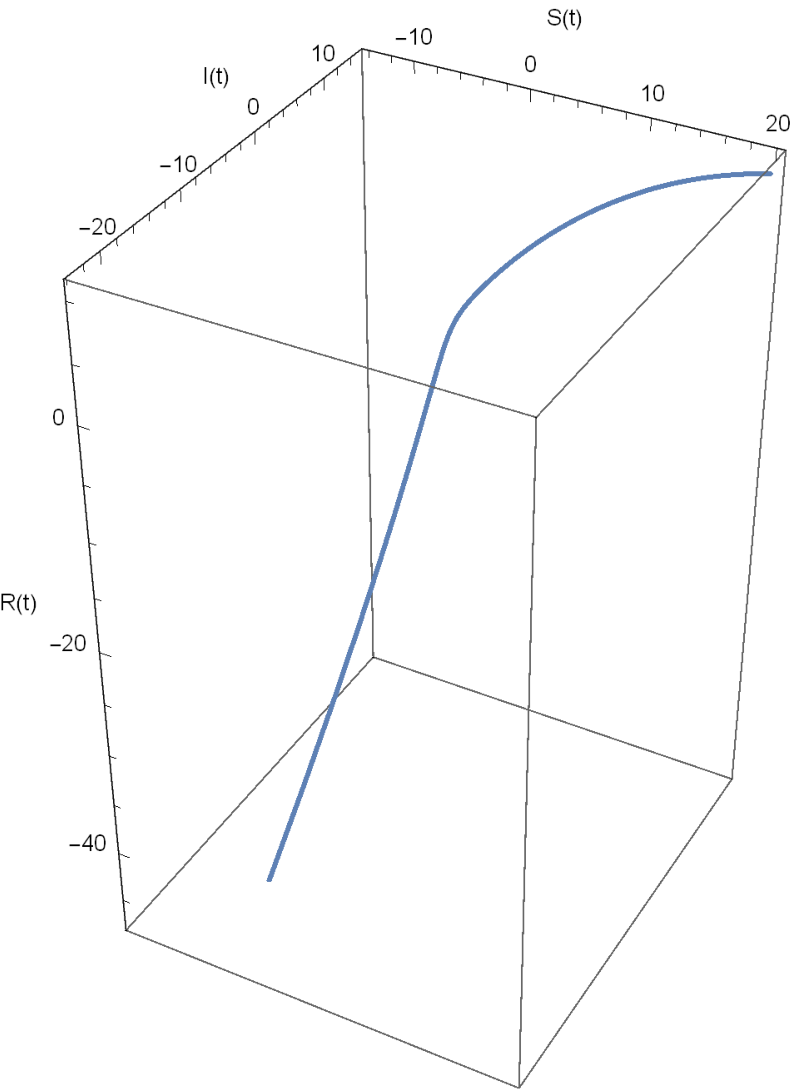}&~~~~~~~~~~~&
 \includegraphics[width=2in]{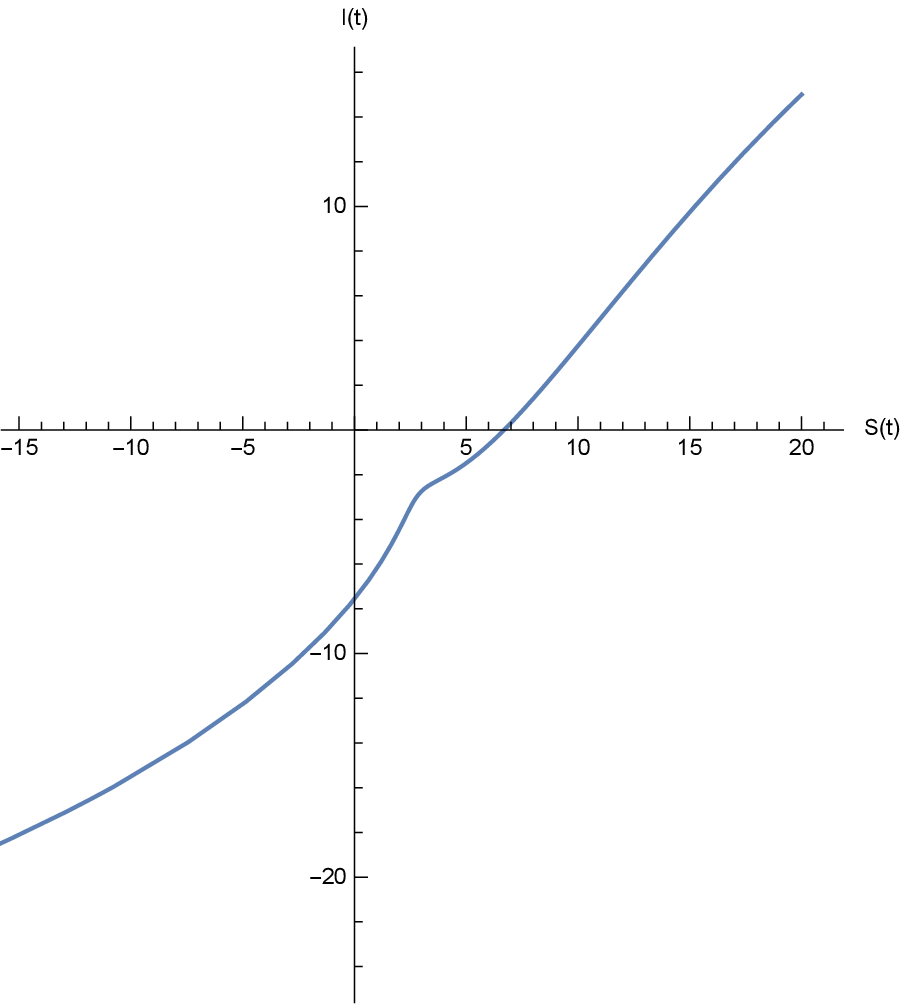}\\
 \\
 \includegraphics[width=2in]{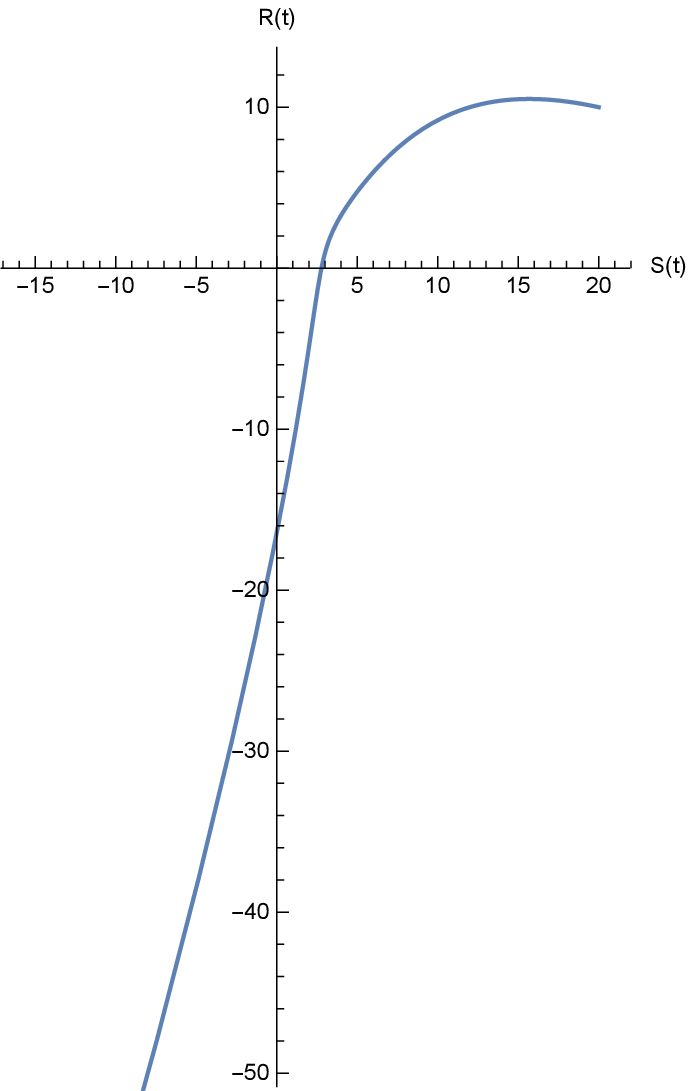}&~~~~~~~~~~~&
 \includegraphics[width=2in]{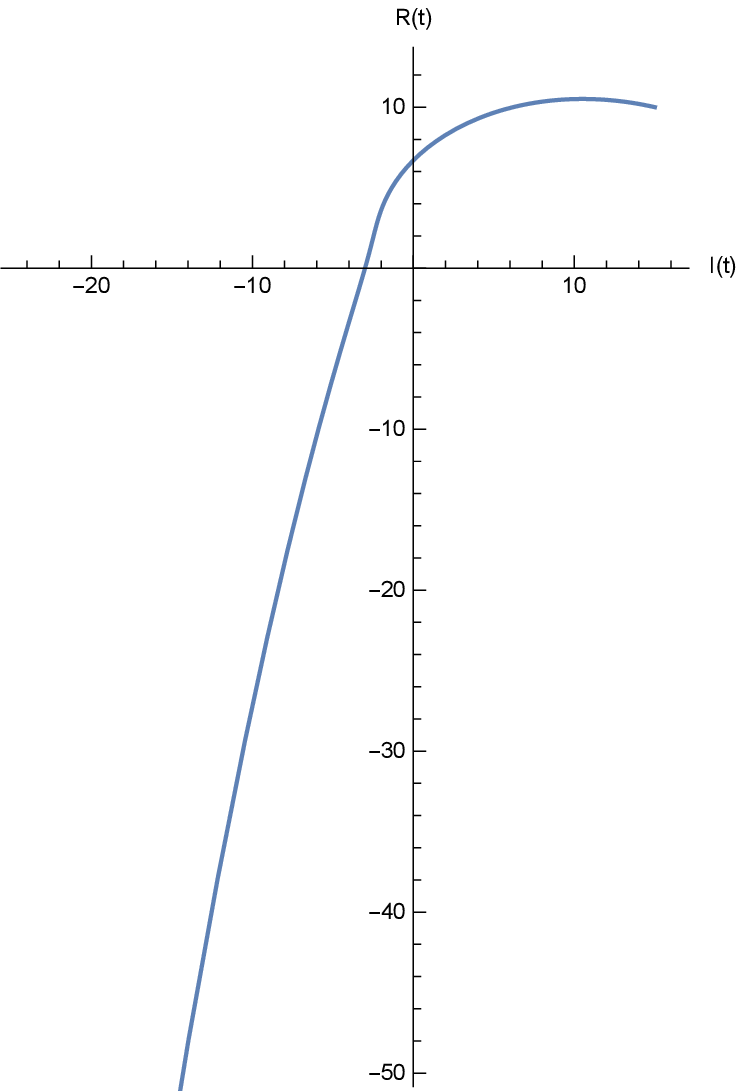}\\
\end{array}$$
  \caption{Phase portraits of $S_{10}(t), I_{10}(t), R_{10}(t)$ by using the LADM. }\label{f4}
\end{figure}

\begin{figure}
\centering
$$\begin{array}{c}
 \includegraphics[width=3in]{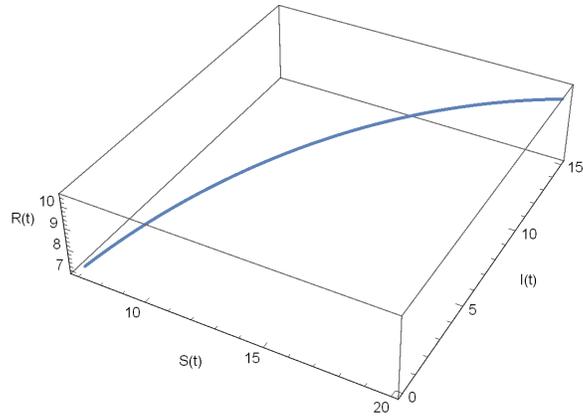}\\
 \\
 \includegraphics[width=2in]{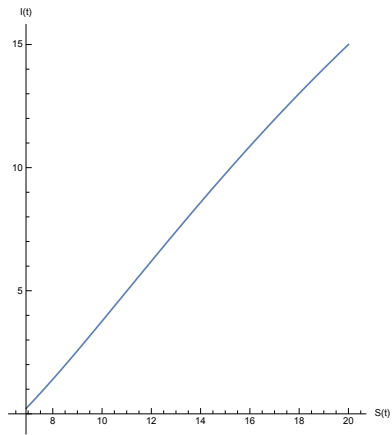}\\
 \\
 \includegraphics[width=3in]{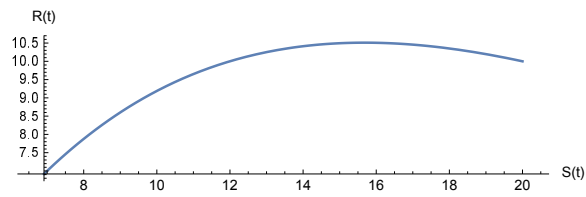}\\
 \\
 \includegraphics[width=3in]{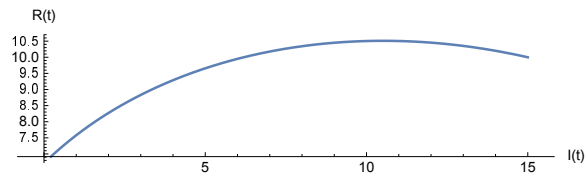}\\
\end{array}$$
  \caption{Phase portraits of $S_{10}(t), I_{10}(t), R_{10}(t)$ by using the DTM. }\label{f5}
\end{figure}

The comparative graphs between the residual errors of the LADM, the
DTM and the HATM for $n=5,10,15$ are demonstrated in Figs. \ref{f1},
\ref{f2} and \ref{f3}. Also, phase portraits of $S-I, S-R, I-R$ and
$S-I-R$ which are obtained by $10$-th order approximation of the DTM
and the LADM are presented in Figs. \ref{f4} and \ref{f5}. According
to the generated results, the LADM has suitable scheme than the DTM
and the HATM.

\begin{table}[h]
\caption{ Numerical comparison of residual error $E_{n,S}(t)$
between LADM, DTM and HATM for $n=5,10$. }\label{t2}
 \centering
\scalebox{0.6}{
\begin{tabular}{|c|c|c|c|c|c|c|}
  \hline
$t$  &$ E_{5,S}(t) $-LADM          &$E_{5,S}(t) $-DTM      &$E_{5,S}(t) $-HATM&        $ E_{10,S}(t)$-LADM             &$ E_{10,S}(t) $-DTM    &$ E_{10,S}(t) $-HATM \\
\hline
$0.0$&$ 0 $                      &$0 $                      &$0$         &$0 $                            &$0 $                     &$0$ \\
$0.2$&$1.98295\times 10^{-11}$   &$6.22316\times 10^{-8}$   &$0.0000116327$        &$4.44089\times 10^{-16}  $       &$4.44089\times 10^{-16}$                       &$4.22713\times 10^{-11}$\\
$0.4$&$4.38413\times 10^{-10}$   &$9.99398\times 10^{-7}$   &$0.0000188901$         &$0 $                              &$1.77636\times 10^{-15}$                     &$8.71449\times 10^{-10}$ \\
$0.6$&$1.87637\times10^{-9}$&$5.07724\times10^{-6}$&$0.0000214386$&$1.77636\times10^{-15}$&$5.9508\times10^{-14}$&$4.95175\times 10^{-9}$\\
$0.8$&$1.93464\times 10^{-9}$  &$0.0000161$                 &$0.00015492$         &$ 2.57572\times 10^{-14} $       &$7.94032\times 10^{-13}$       &$1.57635\times 10^{-8}$\\
$1.0$&$1.18760\times 10^{-8}$  &$0.0000394305$              &$0.000429509$           &$2.30038\times 10^{-13} $       &$5.97122\times 10^{-12}$      &$3.43742\times 10^{-8}$\\
 \hline
 \end{tabular}
 }
\end{table}

\begin{table}[h]
\caption{ Numerical comparison of residual error $E_{n,I}(t)$
between LADM, DTM and HATM for $n=5,10$. }\label{t3}
 \centering
\scalebox{0.6}{
\begin{tabular}{|c|c|c|c|c|c|c|}
  \hline
$t$  &$ E_{5,I}(t) $-LADM   &$E_{5,I}(t) $-DTM   &$E_{5,I}(t) $-HATM &$ E_{10,I}(t)$-LADM &$ E_{10,I}(t) $-DTM    &$E_{10,I}(t) $-HATM  \\
\hline
$0.0$&$0 $                     &$0 $                     &$7.10543\times 10^{-15}$  &$0 $                         &$0 $          &$0$ \\
$0.2$&$2.08857\times 10^{-10}$ &$7.88132\times 10^{-8} $ &$0.000013859$  &$ 4.44089\times 10^{-16}$     &$4.44089\times 10^{-16}$ &$4.24451\times 10^{-11}$ \\
$0.4$&$6.48732\times 10^{-9} $ &$1.26470\times 10^{-6} $ &$0.000036861$  &$4.44089\times 10^{-16} $     &$1.77636\times 10^{-15}$ &$8.8009\times 10^{-10}$ \\
$0.6$&$4.78103\times 10^{-8} $ &$ 6.42035\times 10^{-6}$ &$0.0000397647$  &$ 1.77636\times 10^{-15} $     &$4.44089\times 10^{-14}$ &$5.06861\times 10^{-9}$ \\
$0.8$&$1.95500\times 10^{-7} $&$0.0000203449 $           &$8.51494\times 10^{-6}$  &$3.15303\times 10^{-14}  $    &$5.96412\times 10^{-13}$  &$1.65033\times 10^{-8}$\\
$1.0$&$5.78838\times 10^{-7} $ &$0.000049794 $           &$0.000140914$  &$2.89546\times 10^{-13}  $     &$4.50395\times 10^{-12}$ &$3.7476\times 10^{-8}$ \\
 \hline
 \end{tabular}
 }
\end{table}

\begin{table}[h]
\caption{ Numerical comparison of residual error $E_{n,R}(t)$
between LADM, DTM and HATM for $n=5,10$. }\label{t4}
 \centering
\scalebox{0.6}{
\begin{tabular}{|c|c|c|c|c|c|c|}
  \hline
$t$  &$ E_{5,R}(t) $-LADM   &$E_{5,R}(t) $-DTM &$E_{5,R}(t) $-HATM &$ E_{10,R}(t)$-LADM &$ E_{10,R}(t) $-DTM &$E_{10,R}(t) $-HATM   \\
\hline
$0.0$&$0 $                       &$0  $                   &$0$  &$0  $                          &$0$            &$0$ \\
$0.2$&$5.69638\times 10^{-10} $  &$6.38387\times 10^{-8}$ &$1.95392\times 10^{-6}$  &$1.11022\times 10^{-16} $  &$3.33067\times 10^{-16} $&$1.04154\times 10^{-12}$\\
$0.4$&$1.82284\times 10^{-8} $  &$1.02142\times 10^{-6}$  &$0.0000285358$ &$3.33067\times 10^{-16} $  &$1.22125\times 10^{-15}  $&$8.17403\times 10^{-12}$\\
$0.6$&$1.38422\times 10^{-7} $  &$5.17094\times 10^{-6}$  &$0.000112891$ &$7.21645\times 10^{-16}$  &$4.56857\times 10^{-14}  $&$1.57476\times 10^{-10}$\\
$0.8$&$ 5.83309\times 10^{-7} $  &$0.0000163427 $         &$0.000290976$  &$9.43690\times 10^{-15} $  &$6.10734\times 10^{-13}  $&$1.69486\times 10^{-9}$\\
$1.0$&$1.78012\times 10^{-6} $   &$0.0000398992 $         &$0.000601621$  &$8.77631\times 10^{-14} $&$4.55164\times 10^{-12}  $&$8.65934\times 10^{-9}$\\
 \hline
 \end{tabular}
 }
\end{table}

~\\
~\\
~\\

\section{Conclusion}
In this study, two robust and applicable methods, the DTM and the
LADM were applied to solve the non-linear epidemiological model of
computer viruses.
 In order to show the efficiency and accuracy of presented method,
 the residual errors for different iterations were presented based on
 the LADM, DTM and HATM. Also, the graphs of residual error were
 demonstrated to show the abilities of the LADM than the
 other methods.

\end{document}